\documentclass[twocolumn,aps,showpacs,amssymb,floatfix]{revtex4}
\newcommand{\be}{\begin{equation}}
\newcommand{\ee}{\end{equation}}
\newcommand{\beq}{\begin{eqnarray}}
\newcommand{\eeq}{\end{eqnarray}}
\usepackage{bm}% bold math
\usepackage{graphicx}%
\usepackage{epsf}
\usepackage{epsfig}

\usepackage{amsmath}
\usepackage{amsfonts,amssymb}

\begin{document}
\title{Evidence for diquarks in lattice QCD}
\author{C.~Alexandrou~$^a$, Ph.~de Forcrand~$^{b,c}$ and B.~Lucini~$^{b,d}$}
\affiliation{{$^a$ Department of Physics, University of Cyprus, CY-1678 
Nicosia, Cyprus}\\
{$^b$ Institute for Theoretical Physics, ETH Zurich, CH-8093 Zurich, Switzerland}\\ 
{$^c$ CERN, Physics Department, TH Unit, CH-1211 Geneva 23, Switzerland}\\
{$^d$ Department of Physics, University of Wales Swansea, SA2 8PP Swansea, UK}
}
%\date{\today}%
\begin{abstract}
Diquarks may play an important role in hadron spectroscopy, baryon decays and
color superconductivity. We investigate the existence of 
diquark correlations in lattice QCD by considering systematically all
the lowest energy diquark channels in a color gauge-invariant setup. 
We measure  
mass differences between the various channels and show that
the positive parity scalar diquark is the lightest.
Quark-quark correlations inside the diquark are clearly seen
in this channel, and yield a diquark size of ${\cal O}(1)$ fm.
\end{abstract}
\maketitle

Diquarks were originally proposed several decades ago~\cite{diquarks}
 as a natural way to
explain the absence of a large number of exotics otherwise
predicted by QCD. Recently,  it has been 
realized that they 
provide a natural explanation for an exotic baryon antidecuplet, the
$\Theta^+$~\cite{JW}, that can not be accomodated in the quark model.
The first evidence for the $\Theta^+$ was reported
by the LEPS Collaboration~\cite{Spring8},
but subsequent experiments cast doubts  on
its existence~\cite{review}.
Independently of whether a pentaquark exists or not,
diquark correlations may play an  important role
in the description of quark distributions and
fragmentation functions and
in explaining the systematics of non-leptonic weak decays of light
quark hadrons~\cite{Jaffe}. In addition diquarks are the central ingredient 
of cold, dense matter where they condense to form a color superconductor.
Despite their potential role in explaining such a variety of
phenomena in hadronic physics, quantitative analyses that can directly
assess whether diquark correlations are present in QCD  are lacking.
Lattice QCD is the method of choice for studying hadronic states.
However, 
diquarks are  colored objects and
  need an appropriate formalism to study them using lattice simulations. 
One way is to fix the gauge as done in an early  study on the mass of  
the diquarks~\cite{Karsch}.  
 In this work we present a gauge invariant formalism, where we
create color singlet states by considering
 diquarks  in
the background of a static quark~\cite{latt05,Orginos}. This enables us to extract
 mass differences between  baryons
containing diquarks in the various channels. These mass differences,
unlike diquark masses themselves, are well-defined and gauge
invariant.

 One gluon exchange indicates quark-quark attraction in the
color antitriplet $\bar{3}_c$, flavor antisymmetric $\bar{3}_f$
  and spin singlet positive
parity channel. Diquarks in this channel are referred to 
as ``good'' diquarks~\cite{Jaffe}. The aim of this work is to check whether QCD dynamics 
supports attraction in this channel 
and compare it with other diquark channels.
 Possible diquark configurations are created by two quark operators and
insertions of the covariant derivative.
In this work we consider only
diquark configurations with no derivatives since these are lower in
energy. We therefore consider all 16 diquark multiplets
that can  be created by operators bilinear in the quark fields
of the form $q^T C\Gamma q$ with $C=i\gamma_0\gamma_2$ the charge conjugation
 operator,
and $\Gamma=1, \>\gamma_{\mu}, \>\gamma_5,\>
\gamma_5\gamma_{\mu}, \> \sigma_{\mu \nu}$.
The positive parity channels are $q^T C\gamma_5 q$ and
$q^T C\gamma_5\gamma_0 q$ with spin zero and 
 $q^T C\gamma_i q$ and $q^T C\sigma_{0i} q$ with spin one.
The negative parity channels are $q^T Cq$  and $q^T C\gamma_0 q$ with spin zero
 and $q^T C\gamma_5\gamma_i q$ and $q^T C\sigma_{ij} q$ with spin one.
They create states that vanish in the non-relativistic
limit and are excluded in quark models. We expect them
to be higher in energy.

In addition to mass differences, we
probe correlations directly by examining the spatial distribution of
the two quarks forming the diquark
 in the background of the static quark~\cite{latt05}. 
Correlations which persist as the static quark is removed are intrinsic
to the diquark.
We study these correlations using
gauge invariant two-density correlators, used before~\cite{AFT}
to probe hadron structure:
\small
\be
C_\Gamma({\bf r}_u,{\bf r}_d, t) = \langle 0|J_\Gamma({\bf 0},2t) 
J_0^u({\bf r}_u,t)J_0^d({\bf r}_d,t) J^\dagger_\Gamma({\bf 0},0)|0 \rangle
\label{density_correlator}
\ee
\normalsize
where $J_0^f({\bf r},t)=:\bar{f}({\bf r},t)\gamma_0 f({\bf r},t):, \>f=u,d$ and
\be
J_\Gamma(x)=\epsilon^{abc}\biggl 
[u{^T}_a(x) C\>\Gamma d_b(x)\pm d{^T}_a(x) C\>\Gamma u_b(x) \biggl ]s_c(x)
\ee
%where the positive (negative) sign corresponds to the flavor symmetric (antisymmetric)
where the + (-) sign corresponds to the flavor symmetric (antisymmetric)
combination, 
and $s_c$ denotes the static quark.
Latin indices denote color.
We use two degenerate flavors of Wilson fermions on quenched
configurations generated at $\beta=5.8$, 6.0 and 6.2 corresponding to 
lattice spacing $a=0.136$, 0.093 and 0.068 fm as determined from the static
quark force~\cite{sommer}. At each lattice spacing, we perform
measurements at three values of the 
pion mass
in the range $570-910$ MeV. 
By comparing results obtained at different lattice spacings
but at the same pion mass  we can assess  
 discretization effects. In most cases the
quenched approximation is good
when the mass of the pion is higher than $600$ MeV, but is
expected to
fail in the chiral limit. In order to check the validity of our results in
this range of pion masses, we repeat our mass splitting measurements on
a set of configurations with two degenerate flavors of dynamical Wilson fermions at
$\beta=5.6$~\cite{newSESAM}. 
In Table~\ref{table:parameters} we collect the parameters of our calculation.

\begin{table}[h]
\caption{Summary of our simulations, including the value of $\kappa$, which determines the
bare quark mass,  the pion mass, $m_\pi$,
and nucleon mass, $M_N$,  in lattice and, for the latter,
 in physical  units and 
 the number of gauge
configurations.}
\begin{tabular}{|c|c|c|c|}
\hline
\multicolumn{1}{ |c|}{$\kappa$  } &
\multicolumn{1}{ c|}{$am_\pi$  } &
\multicolumn{1}{ c|}{$aM_N$ [$M_N$~GeV]} &
\multicolumn{1}{|c|}{number of confs } 
\\
\hline
\multicolumn{4}{|c|}{Quenched $16^3\times 32$ \hspace*{0.2cm}$\beta=5.8$, $a^{-1}=1.47$~GeV }
 \\ 
    0.1560 &  0.619(2) & 1.139(8) [2.04(1)]&  356  \\
    0.1575 &  0.549(2) & 1.052(9) [1.55(1)]&  160 \\
    0.1590 &  0.473(2) & 0.961(9) [1.41(1)]& 200 \\
\multicolumn{4}{|c|}{Quenched $16^3\times 32$ \hspace*{0.2cm}$\beta=6.0$, $a^{-1}=2.15$~GeV }\\
   0.153   & 0.423(1) & 0.783(8)[1.68(2)] &  364 \\
   0.154   & 0.366(2) & 0.716(7)[1.54(2)] & 503 \\
   0.155   & 0.300(2) & 0.634(8)[1.36(2)] & 287 \\
\multicolumn{4}{|c|}{Quenched $20^3\times 40$  \hspace*{0.2cm}$\beta=6.2$, $a^{-1}=2.94$~GeV }
\\ 
 0.1510 & 0.286(3) & 0.563(8) [1.66(2)]& 116 \\
 0.1520 & 0.214(3) & 0.487(9) [1.43(3)]& 166 \\
 0.1523 & 0.188(3) & 0.460(11)[1.35(3)] & 157 \\
\multicolumn{4}{|c|}{Unquenched Wilson $24^3\times 40 $ \hspace*{0.2cm}$\beta=5.6$, $a^{-1}=2.42$~GeV }
 \\
 0.1575  & 0.270(3) & 0.580(7)[1.40(2)] & 185 \\
\hline
\end{tabular}
\label{table:parameters}
\vspace*{-0.3cm}
\end{table} 

The mass of our static-light-light baryon 
can be obtained from the large time limit of the 
 correlator 
$G_\Gamma(t)=\langle J_\Gamma(\vec{x},t) J_\Gamma^{\dagger}(\vec{x},0)
\rangle$.
The effective mass, 
$m_{\rm eff}(t)\equiv-\log\left[G_\Gamma(t)/G_\Gamma(t-1)\right]$,
becomes time independent for large $t$ when the lightest state dominates 
(plateau region),
yielding its mass. 
To reduce the
 statistical errors,
we need to keep $t$ small 
by isolating the ground state  as fast as possible.
 This is accomplished by suppressing higher excitations using Wuppertal 
smearing~\cite{Wuppertal}
on the source and sink with hypercubic (HYP) smeared spatial
links~\cite{Hasenfratz:2001hp} for the Wuppertal smearing function~\cite{NN}.
 In addition
we  use HYP smearing on the temporal gauge links $U_4(x)$
  that enter in the construction
of the static propagator given by
\small
\be
S_{\rm stat.}({\bf x}_2,t_2,{\bf x}_1,t_1)= \delta^3({\bf x}_2-{\bf x}_1)\biggl(\frac{1+\gamma_0}{2}\biggr) \left[\prod_{t=t_1}^{t_2-a} U_4({\bf x}_1,t)\right]^\dagger
\label{static prop.}
\ee
\normalsize
for $t_2>t_1$. 
The exponential prefactor $\exp(-m_q(t_2-t_1))$ has been dropped, since this
introduces a constant shift in all energies by the bare heavy quark mass $m_q$,
which cancels in the mass differences that we measure between baryons with
diquarks in different channels.

\begin{figure}[h]
\epsfxsize=8truecm
\epsfysize=5.5truecm
\mbox{\epsfbox{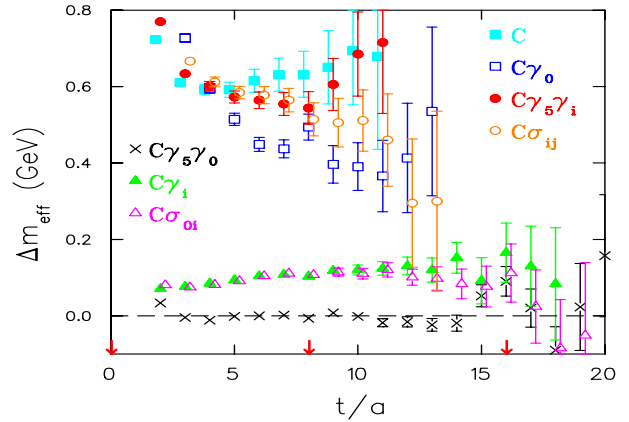}}
\caption{Effective mass difference, 
$\Delta m_{\rm eff}$, in the various diquark channels,
 at $\beta=6.0$ for our lightest quarks. % ($\kappa=0.155$). 
%Crosses correspond to taking $\Gamma=\gamma_5\gamma_0$, filled triangles
% $\Gamma=\gamma_i$, open triangles  $\Gamma=\sigma_{0i}$,
%filled squares  $\Gamma=I$, open squares  $\Gamma=\gamma_0$, filled circles 
% $\Gamma=\gamma_5 \gamma_i$ and open circles to $\Gamma=\sigma_{ij}$.
The arrows (from left to right) indicate the time slice of the source, 
of the density insertions and of the sink.}
\label{fig:diquark_mass_difference}
\vspace*{-0.25cm}
\end{figure}

%In Ref.\cite{Jaffe} it is argued that the ``good'' (scalar) diquark 
%is lower in energy
%than the ``bad'' (vector) diquark by about $2/3$ of the mass splitting
%between the nucleon and the $\Delta$ i.e. by about 200~MeV.
We show in Fig.~\ref{fig:diquark_mass_difference} the effective
mass differences, $\Delta m_{\rm eff}$, between the ``good'' (scalar)   
diquark
created by taking
 $\Gamma= \gamma_5 $, and the
other channels. As expected  the mass difference is zero for the other
scalar diquark with positive parity created by taking 
$\Gamma=\gamma_0 \gamma_5 $. 
The other positive parity channels, i.e. the ``bad'' (vector) diquarks,
 are degenerate and are
clearly higher in energy than the ``good'' diquarks as pointed out 
in Ref.~\cite{Jaffe}.
The negative parity 
states, as expected, have even larger energies.
Their effective masses  are much noisier making it
difficult to identify a clear plateau void of contamination  
from yet higher states. The general behavior is however similar to the
positive parity diquarks: the vector channels are also degenerate 
and tend to have a larger mass than
the scalar channel created by taking $\Gamma=I.$ 
To verify that the  mass difference seen between the two
scalar channels
is significant  one would need
to improve the
quality of the mass plateaus.

\begin{figure}[h]
\epsfxsize=8truecm
\epsfysize=5.truecm
\mbox{\epsfbox{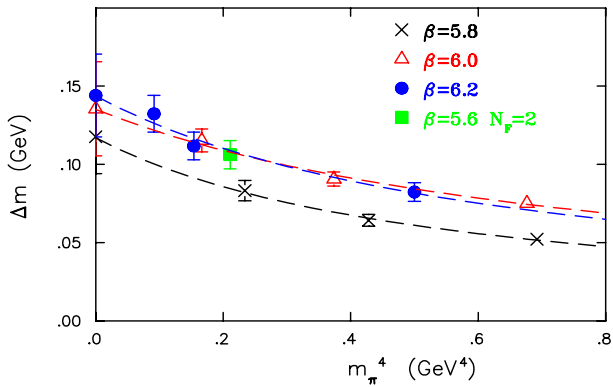}}
\caption{Mass difference between ``good'' and ``bad'' diquarks
as a function of $m_\pi^4$ at $\beta=5.8$ (crosses), $\beta=6.0$ 
(open triangles), $\beta=6.2$ (filled circles) and unquenched result 
(filled square). The dashed lines are fits to Eq.~(\ref{eq:m_continuous}).}
\label{fig:dm}
\vspace*{-0.25cm}
\end{figure}

Effective color-spin Hamiltonian arguments~\cite{Jaffe} predict that for heavy
constituent quarks the mass difference $\Delta m\equiv m_{\gamma_i}- m_{\gamma_5}$ between the ``good'' and ``bad''
diquarks scales like 1/$m_{q_1} m_{q_2}$, where $m_{q_{1,2}}$ are the
masses of the constituent quarks.
On the other hand in the light quark regime we have $ m_\pi^2 \propto m_q$.
Therefore for intermediate values
of $m_q$ where both relations hold approximately  we  make the Ansatz
  $\Delta m\propto 1/m_{\pi}^4$. In the
opposite limit $m_q \to 0$, $\Delta m$ is expected to
 approach
a constant value. These two behaviors 
can be connected by the Ansatz
\beq
\label{eq:m_continuous}
\Delta m = \frac{c_1}{1 + \left(m_{\pi}/c_2\right)^4},
\label{mass dependence}
\eeq
with $c_1$ and $c_2$ to be determined from the lattice data.
We show in Fig.~\ref{fig:dm} results for $\Delta m$ as a function of $m_\pi^4$
for our three $\beta$ values. All lattice data should fall on a universal line
if we are close to the continuum limit. We observe that this is indeed the
case on our two finest lattices 
whereas for the coarsest, scaling violations are apparent.
In the same figure we also show $\Delta m$ obtained
using unquenched, dynamical Wilson configurations. It nicely falls on the 
same curve as the fine-lattice quenched results.
This corroborates that scaling has set in and that quenching effects
at these quark masses are small.  Therefore a quenched study
of diquark properties in the quark  mass range used here is a very good
approximation and will be adopted in the rest of the discussion.
The Ansatz given in Eq.~(\ref{eq:m_continuous}) provides a good fit
to the lattice data.  Using the lattice data with good
scaling behavior at $\beta=6.0$ and $6.2$, 
 we find $c_2=0.78\pm 0.15$~GeV, and $c_1=0.138\pm 0.01$~GeV, which is the mass
difference at the chiral limit~\cite{Kostas}.
This mass splitting can be compared to the  $\Delta$-nucleon mass splitting, 
$\delta m_{\Delta N}$. 
Using our quenched data at $\beta=6.0$ we find for
the ratio  $\Delta m/\delta m_{\Delta N}=0.67(7),\> 0.73(8)$ and $0.67(8)$ 
at $\kappa=0.153,\>0.154$ and $0.155$ respectively.
Therefore our
estimate of $\Delta m$
 is consistent with the predicted $2/3$ of the $\Delta$-nucleon mass 
difference given in Ref.~\cite{Jaffe}.

\begin{figure}[h]
\epsfxsize=8truecm
\epsfysize=6truecm
\mbox{\epsfbox{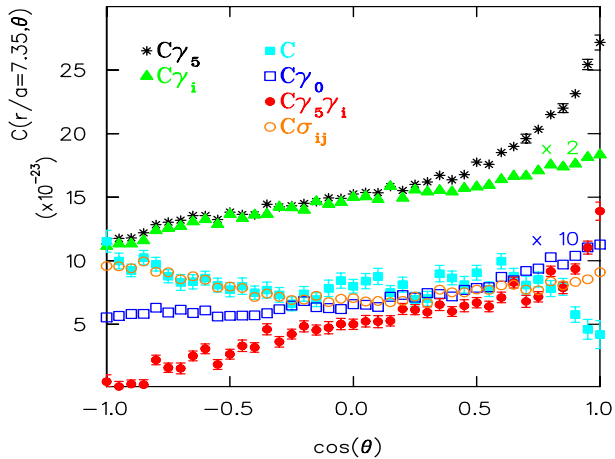}}
\vspace*{-0.25cm}
\caption{$C_\Gamma(r/a=7.35,\theta)$ versus cos$(\theta)$ for
$\beta=6.2$ and $\kappa=0.1520$
 for the  ``good'' diquark 
( asterisks) and the other diquark channels
using the same
notation is as in Fig.~\ref{fig:diquark_mass_difference}.
Correlators for the ``bad'' diquark have been multiplied by 
two and for all negative parity channels by ten.}
\label{fig:density_beta_6.2}
\end{figure}
\begin{figure}[h]
\epsfxsize=8truecm
\epsfysize=10truecm
\mbox{\epsfbox{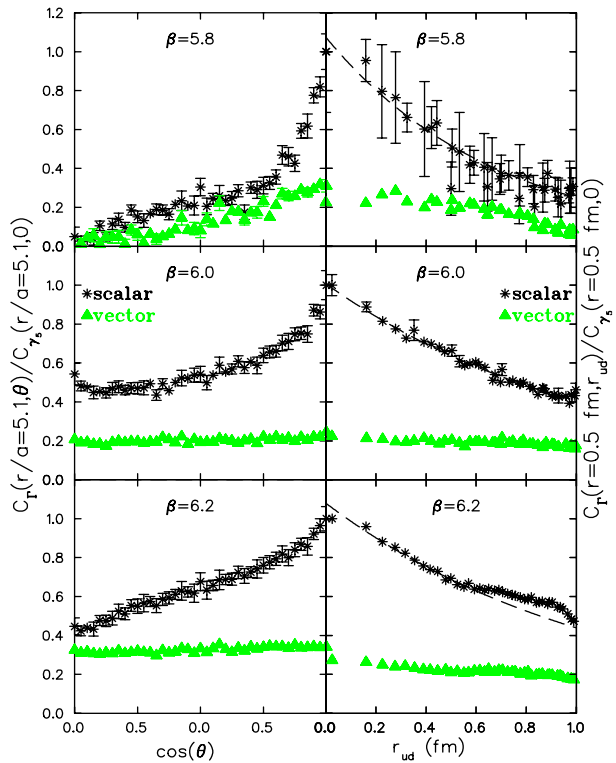}}
\caption{Left: $C_{\Gamma}(r/a=5.1,\theta)/C_{\gamma_5}(r/a=5.1,0)$ versus cos($\theta$). Right:
$C_{\Gamma}(r=0.5\>{\rm fm},r_{ud})/C_{\gamma_5}(r=0.5 \>{\rm fm},0)$ 
versus  $r_{ud}$, 
 for  the ``good''  (asterisks) 
and  ``bad''  (filled triangles)  diquarks at the lightest pion  for
our three lattice spacings.}
\label{fig:all_betas}
\vspace*{-0.5cm}
\end{figure}

Having determined the diquark spectrum, 
we now turn to the analysis of their
structure
by studying the density-density correlators defined in
Eq.~(\ref{density_correlator}).
The time $t$ where the density operators are inserted is shown in
Fig.~\ref{fig:diquark_mass_difference} by the arrow (at $t/a=8$),
and is within the plateau range
of the effective mass. We have verified that the correlators 
remain, within errors, unchanged when we vary the source-sink separation
or equivalently $t$.
Our aim is to look for spatial correlations between the two light quarks
in the various diquark channels. We take  the location
of the static quark  as the reference coordinate from which the
distances ${\bf r}_u$ and ${\bf r}_d$ of the two quarks of
different flavors are measured.
We are interested in intrinsic diquark correlations, which persist as
the static quark is moved away from the diquark. 
We therefore consider spherical shells $|{\bf r}_u|=|{\bf r}_d|=r$ 
of increasing radius $r$.
Since the system is spherically
symmetric the correlator  depends only on $r$ and the angle 
$\theta=\arccos(\hat{\bf r}_u . \hat{\bf r}_d)$~\cite{latt05}.
In the absence of any correlation, the distribution of $C_\Gamma(r,\theta)$
will be uniform as a function of $\cos(\theta)$.
Attraction will show up as an enhancement at small angles, near $\cos(\theta)=1$.
Our cubic lattice breaks rotational symmetry 
and distorts the uniform spherical distribution, particularly at
small angle $\theta$. To remove such lattice artifacts, we normalize 
our distributions by a uniform lattice distribution.
We show the resulting density correlators in Figs.~\ref{fig:density_beta_6.2} and
\ref{fig:all_betas} for various shell radii $r$.  
The physically relevant correlations are those that survive when
$r$ is large. We indeed observe  
that the  qualitative behavior of the distributions 
does not depend on $r$,
 confirming that the color field
generated by the static quark does not affect the physics of diquarks once
$r$ is large enough.

In Fig.~\ref{fig:density_beta_6.2} 
we show $C_\Gamma(r,\theta)$ for all the different channels as a function
of $\cos(\theta)$ when the shell radius is fixed to $r=0.5$~fm ($r/a=7.35$).
 We clearly observe that the ``good'' diquark shows
enhanced correlations at small $\theta$, indicating
attraction between the quarks. There
is also a gradual increase in the distribution for 
the ``bad'' diquark  indicating a weaker attraction
in this channel. This behavior persists
as we increase or decrease the quark masses.
For the negative parity channels, correlations are absent or very weak,
except for the vector channel with $\Gamma=\gamma_5 \gamma_i$.
However those disappear for lighter quark masses.
The negative-parity correlators are noisier because the diquarks are heavier,
so that a higher statistics analysis will be needed to fully
determine  the characteristics of their distributions.
 For this reason, 
we now restrict our analysis to the positive parity diquarks.
\begin{figure}[h]
\hspace*{-2.8cm}
\scalebox{0.4}{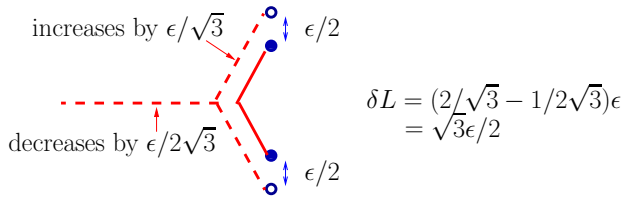}
\caption{String formation for three quarks, where one quark has been taken
to the distant left. As a function of the separation between the other two quarks, 
the static potential rises with an effective string tension 
$\sigma_{\rm eff} = \sigma \frac{\sqrt{3}}{2}$.}
\label{fig:size}
\vspace*{-0.25cm}
\end{figure}

In Fig.~\ref{fig:all_betas} we show the density correlators for the
``good'' (scalar) and ``bad'' (vector) diquarks for all three lattice spacings
at our lightest quark mass.
The left hand set shows the correlator as a function of $\cos(\theta)$   normalized to 1 
at $\theta=0$ for $r/a=5.1$, which
corresponds to  $r=0.69$, 0.47 and 0.35~fm at $\beta=5.8$, 6.0 and 6.2
respectively. As can be seen, irrespectively of the distance the ``good'' 
diquark shows stronger spatial correlations. Within this
framework we can also extract the diquark size: At fixed $r$ we
look at the dependence of the correlator on the relative $u-d$ separation, 
$r_{ud}=2r\sin\left(\theta/2\right)$.
In the right hand of Fig.~\ref{fig:all_betas},
we show the correlator (normalized to 1 at the origin)
as a function of $r_{ud}$ for a fixed physical
shell radius $r=0.5$~fm, at our lightest quark mass.  
 An exponential dependence of 
$C_{\gamma_5}(r,r_{ud})\propto \exp(-r_{ud}/r_0(r))$ provides a gauge invariant
definition of the diquark size $r_0(r)$ for a 
given value of $r$. The curves in Fig.~\ref{fig:all_betas},
 obtained from fits to an exponential dependence, describe well
 $C_{\gamma_5}(r,r_{ud})$
at all $\beta$ values. 
Moreover the physical size $r_0(r)$ is the same on the two finer lattices,
confirming continuum-like behaviour as for the diquark masses.
We  can then evaluate $r_0$ as a function of the shell 
radius $r$. 
Our measurements show a mild increase of $r_0$ with $r$, from $\sim 0.9$~fm
to $\sim 1.3$~fm as $r$ increases from $0.3$ to $0.75$~fm.
For large $r \gtrsim 0.5$~fm, $r_0$ is consistent with having reached a plateau.
 This suggests that the static source has
a small influence on the ``good'' diquark giving a
characteristic size of about $1.1\pm 0.2$~fm.
For comparison, 
 at the same quark mass and using the same
definition of size, we find for the $\rho$ a size of 0.7~fm~\cite{AFT}. 
The large diquark size can be understood from the following qualitative argument:
As the static quark is moved away from the diquark a $q$-$q$ string
 tension develops. Considering an increase $\epsilon$ in the separation
between the quarks in the diquark as illustrated in Fig.~\ref{fig:size}
one finds that the  effective $q$-$q$ string
 tension
 is $\left(\sqrt{3}/2\right)\sigma$ where $\sigma$ is the $q$-$\bar {q}$
string tension.
Since  in the perturbative regime, the $q$-$q$ attraction is also weaker
than the $q$-$\bar{q}$ attraction, this time by a factor $1/2$,
the conclusion is that for all distances one expects
the $q$-$q$ attraction to be weaker. Thus a $q$-$q$ diquark should be somewhat larger than
a  $q$-$\bar{q}$ meson, which is what we find.
Size measurements for  the ``bad'' diquark, on the other hand, show neither
scaling nor convergence to a plateau value. The large
values obtained, often similar to our box size, corroborate the weakness of spatial
correlations in this channel.

In conclusion, we have evaluated the mass splittings and density correlators of
the complete set of diquark channels created by local diquark
fields.
Both observables confirm the phenomenological
expectation that QCD dynamics favors the formation 
of ``good'' diquarks, i.e. in the scalar positive parity channel. The characteristic
size of this diquark, ${\cal O}(1)$~fm, is  large 
but consistent with the scale  ${\cal O}(200)$ MeV of the attraction.
Even a ``good'' diquark is a large object, which may limit its 
relevance to hadron structure.
The  positive parity vector channel is higher in energy by about $2/3$ the $\Delta$-nucleon
mass splitting, and forms an even larger object.
All the negative parity channels
have much higher energies.

{\bf Acknowledgments:}
We  thank B. Orth,  Th. Lippert and K. Schilling~\cite{newSESAM}
for providing the unquenched configurations.
This work is
supported in part by the  EU Integrated Infrastructure Initiative
Hadron Physics (I3HP) under contract RII3-CT-2004-506078.
The work of B.L. has been
partially supported by the Royal Society.


\begin{thebibliography}{99}
\bibitem{Kostas} Diquark mass splittings, measured in Ref.~\cite{Orginos} for lighter quarks  tend to be larger than ours, with
large uncertainties.
\end{thebibliography}
\end{document}